\documentclass[11pt]{article}
\usepackage{hfstyle}
\usepackage[utf8]{inputenc}

\usepackage{amsthm, amsmath,graphicx,mathabx,bbm, color}
\usepackage[normalem]{ulem}

\newtheorem{proposition}{Proposition}
\newtheorem{conjecture}{Conjecture}

\newcommand{\ZZ}{{\mathbbm{Z}}}
\newcommand{\NN}{{\mathbbm{N}}}

\newcommand{\card}{\mathop{\mathrm{card}}}

\newcommand{\supp}{\mathop{\mathrm{supp}}}
\newcommand{\diam}{\mathop{\mathrm{diam}}}

\begin{document}
\title{Deterministic cellular automata resembling  diffusion}
\author{Henryk Fuk\'s and Sanchala Abeykoon Mudiyanselage 
      \oneaddress{
         Department of Mathematics and Statistics\\
         Brock University\\
         \email{hfuks@brocku.ca, sa18hz@brocku.ca}
       }
   }

\Abstract{
We investigate number conserving cellular automata with up to five inputs
and two states 
with the goal of comparing their dynamics with  diffusion. For this purpose, we introduce the concept
of decompression ratio describing
expansion of configurations with finite support. We find that a large number of
number-conserving rules exhibit
abrupt change in the decompression ratio
when the density of the initial pattern 
is increasing, somewhat analogous to the second order phase transition.
The existence of this transition is formally
proved for rule 184.
Small number of rules exhibit
infinite decompression ratio, and such rules may be useful for ``engineering''
of CA rules which are good models of
diffusion, although they will most likely require more than two states.
}
\maketitle


\section{Introduction}
The phenomenon of diffusion is very common in various natural, technological, economic and social systems.
What is usually understood by it is the movement of particles (or ideas, biological organisms, people, price values, etc.)  from a region of higher concentration to a region of lower concentration. At the microscopic level, the driver of the diffusion is the random walk or Brownian motion of individual particles.
When diffusion is simulated in computer software,  for example, in order to numerically investigate a model of a natural phenomenon, random
walks must be simulated using pseudo-random number generator, consuming significant CPU resources. Alternatively, diffusion can be
modeled and studied using partial differential equations,  such as the heat equation or Fick's laws of diffusion. Such equations must be
first discretized in order to solve them numerically on digital computers.

If the objective is to construct a computer model which exhibits movement form higher to lower concentrations,
it would be obviously best to avoid using random numbers and differential equations, devising the model from the beginning in such a way that it operates in the space of binary strings.
All what is then needed is to find the right dynamical system ``living'' in such a space and resembling diffusion to the extent it is possible.
Cellular automata seem to be ideally suited for this purpose, since they are fully discrete analogs of partial differential equations, 
with discrete space, time, and state variables.

Our goal in this paper is, therefore, to search for binary cellular automata which share some features with broadly understood diffusion processes,
and to find out how well they resemble the actual diffusion (if alt all).

In order to have some sort of benchmark to compare with, let us
consider a model of ``real'' diffusion constructed as follows.
Consider one dimensional lattice with lattice sites being either empty (state 0)
or occupied by a single particle (state 1). All particles simultaneously
and independently of each other decide whether to move to the left or to the
right, with the same probability 0.5 in either direction. We then simultaneously move every particle to the desired position if it is empty, otherwise the particle stays in the same place. If two particles want to move to the same empty spot, only one of them, randomly selected, is allowed to do so.
This process, which constitutes a single time step, is then repeated for
as many time steps as desired.

\begin{figure}
 \begin{center}
  \includegraphics[width=12cm]{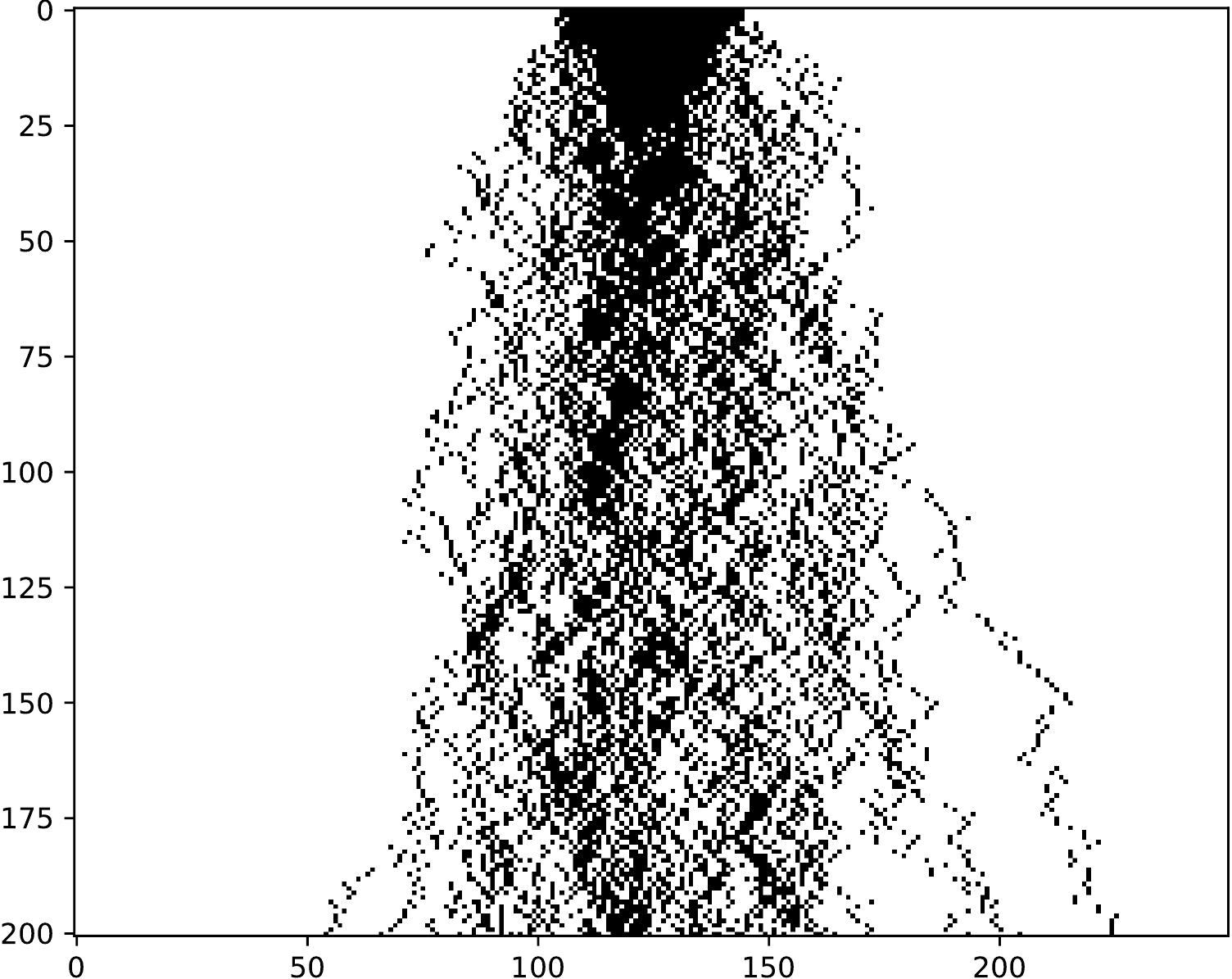}
 \end{center}
\caption{Spatiotemporal pattern of random diffusion on
one dimensional lattice of 250 sites. The initial condition is a solid block of 40 particles. Every 10th time step is shown, so that the numbers on the vertical axis
correspond to the actual time divided by 10.}\label{randif}
\end{figure}

Figure~\ref{randif} illustrates this process for 40 particles initially
occupying a block of 40 sites in the middle of the lattice of 
250  sites.
Consecutive iterations are shown every 10 steps as individual rows.
Black squares represent particles, empty spaces are shown in white color. 
We can see that with the passage of time the particles occupy wider and wider region. This is
analogous to the behaviour of a gas which, if released in a small region of a container, will eventually spread over the entire container.

Three features of the above process are crucial: (i) the ``gas'' of particles expands in space, (ii) the total number of particles is conserved
and (iii) the arrangement of particles appears more and more
disordered (spatial entropy increases). While our ultimate 
goal is to find (or construct) CA rules satisfying all three conditions, in this paper  we will pursue a more modest goal, namely we will investigate binary cellular automata satisfying the first two conditions only. 

\section{Basic definitions}
Let $f: \{0,1\}^n \to \{0,1\}$ be called a \emph{local function} or \emph{rule} of cellular automaton, where $n>0$. We will sometimes refer to it as $n$-input rule.
Let $l$ and $r$ be two positive integers
called, respectively, \emph{left radius} and \emph{right radius} such that 
$l+r+1=n$.
For a given local function $f$, we define the \emph{global function} $F: \{0,1\}^\ZZ
\to \{0,1\}^\ZZ$ such that for all $x \in \{0,1\}$ we have
$$[F (x)]_i = f(x_{i-l},x_{i-l+1}, \ldots,  x_{i+r}).$$
It is customary to refer to CA rules by their Wolfram number \cite{Wolfram94},
defined as 
$$W(f)=\sum_{(x_1,x_2, \ldots x_n)\in \{0,1\}^n} 2 ^{2^{n-1}x_1+2^{n-2}x_2+\ldots2^0 x_n}.$$

When discussing family of specific CA rules it is usually convenient to divide them into \emph{equivalence classes} of rules having similar properties \cite{Wolfram94,b93}. These equivalence classes are defined with respect to the group generated by the operators of \emph{reflection} and \emph{conjugation} denoted, respectively by $R$ and $C$, and defined on the set of all one-dimensional binary $n$-input CA rules $f$ by
\begin{align*}
R\,f(x_1,x_2,\ldots,x_n) & = f(x_n,x_{n-1},\ldots,x_1),\\
C\,f(x_1,x_2,\ldots,x_n) & = f(1-x_1, 1-x_2,\ldots,1-x_n).
\end{align*}
In some cases, equivalence classes generated by the operator $R$ alone are more appropriate, and we will
discuss this  later in the paper.

A one-dimensional  $n$-input CA rule $f$
is \textit{number-conserving} if, for all $L\ge n$ and all 
$(x_1,x_2,\ldots , x_L)\in \{0,1\}^L$
it satisfies
\begin{equation}
\sum_{i=0}^{L-1} f(x_i,x_{i+1},\ldots,x_{i+n-1})
=\sum_{i=0}^{L-1} x_i
\label{CRf}
\end{equation}
where all indices are taken modulo $L$.

Property of being number-conserving is decidable and  easily checked, using the well known method described in \cite{paper12}. These rules have been 
exnensively studied in the past \cite{Hattori91,Formenti2003,Durand2003,Pivato02,Moreira03}.
It is well known that
there exist
\begin{itemize}
\item 5 number conserving rules with 3 inputs, rules 184, 226, 170, 240 and 204, although the local functions of the last three effectively depend on 1 input only;
\item 22 number conserving rules with 4 inputs, although 8 of them depend effectively on 3 or less inputs only.
\item 428 number conserving rules with 5 inputs, 
\end{itemize}

\section*{Decompression ratio for cellular automata}
As remarked in the introduction, we will be searching for rules which 
behave like ``expanding gas'' and which conserve the number of particles. The second condition will obviously be satisfied by number conserving rules, thus we will restrict our attention to such rules from now on. As for the first condition, 
we need to quantify the notion of ``expansion''.

Let $x \in {\{0,1\}}^\ZZ$ be a bi-infinite binary sequence, to be called configuration.
The set of indices $i$ corresponding to nonzero values of $x_i$ will be called support of $x$,
$$\supp x= \{i \in \ZZ: x_i \neq 0\}.$$
If $\supp x$ is finite, then we can define \emph{diameter} of $x$,
$$\diam(x) =\max \supp x - \min \supp x +1$$

For $x$ with finite support, we define \emph{density of support} 
as 
$$ \rho(x) = \frac{\sum_{i \in \ZZ} x_i}{\diam(x)}$$

Now let us suppose that $F$ is the global function of some cellular automaton, so that $F^k(x)$ denotes the configuration obtained by iterating $F$ $k$ times. For $x$ with finite support and for a number-conserving $F$, 
$$\sum_{i \in \ZZ} x_i = \sum_{i \in \ZZ} [F^k(x)]_i$$
for any $k \in \NN$.

Define \emph{decompression ratio} for a given cellular automaton $F$ and a 
configuration $x$  as 
$$ \psi_F(x)= \lim_{k \to \infty} \frac{\diam\left(F^k(x)\right)}{\diam(x)}. $$

\begin{figure}
 \begin{center}
  \includegraphics[width=13cm]{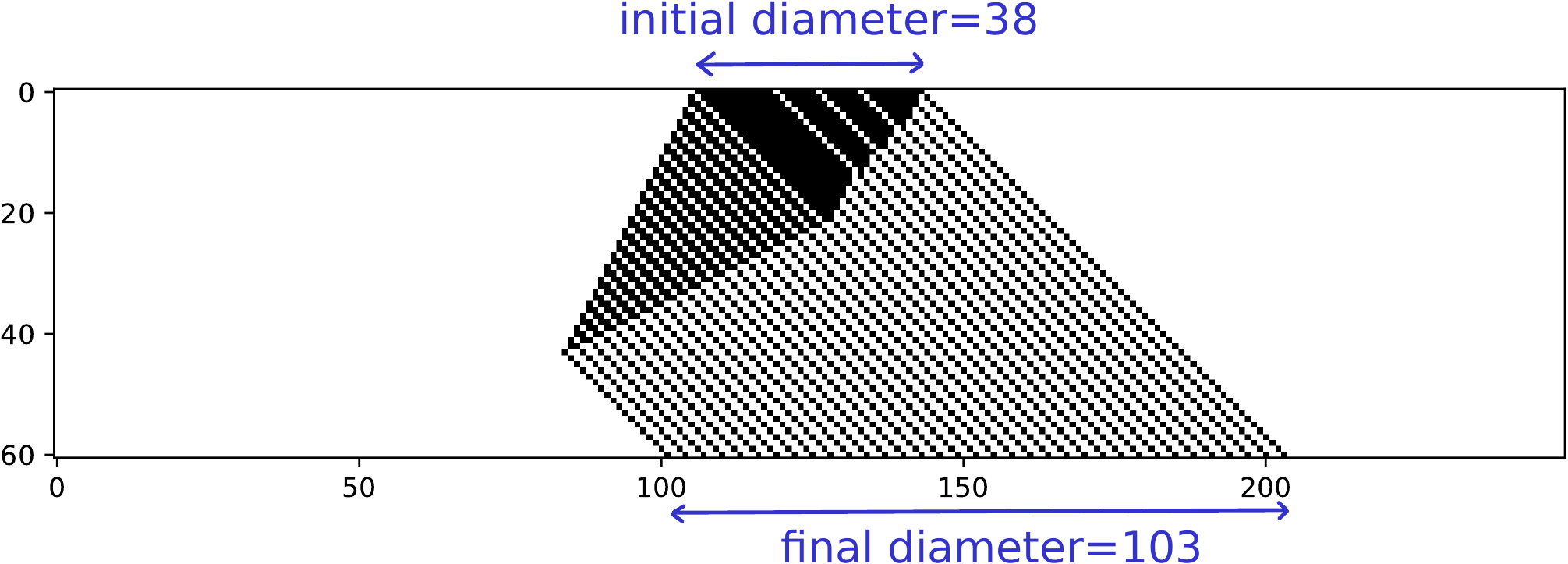}
  \end{center}
  \caption{Calculations of the decompression ratio
  for a sample initial configuration and 4-input rule 60200. Here $ \psi_F(x) = \frac{103}{38}=\approx 2.71$  
  }\label{figex}
 \end{figure}

Figure \ref{figex} shows an   example calculation of the decompression ratio in a concrete situation. In this example the the decompression ratio reaches certain
limit after a finite number of iterations and stays constant thereafter.
As we shall see, this will be a common feature in number-conserving CA.
Since the decompression ratio for a given rule depends on the initial configuration, 
we need to average it out over all configurations. 
We will achieve this by considering the set of all configurations 
which have a given fixed diameter and a given number of ones.
More formally,
let $d>2$, $2 \leq n \leq d$  and and let $D_{d,n}$ be the set of all configurations $x \in {\{0,1\}}^\ZZ$ with support starting at the origin, such that they have 
 exactly $m$ 1's and $\diam x=d$. 
Since the number of binary strings of length $a$
having exactly $b$ 1's is equal to $\binom{a}{b}$,
it immediately follows that
$$\card D_{d,n}= \binom{d-2}{n-2}.$$

Then we define \emph{expected decompression ratio} as
$$ \psi_{d,p,F}=
\frac{\sum_{x \in D_{d,n}} \psi_F (x)}{\card D_{d,n}},
$$
 where $p=n/d$ represents the density of support of configurations
 in $D_{d,n}$. The expected decompression ratio can still possibly depend on 
 the diameter $d$, thus we will define 
 $$ \psi_{p,F}= \lim_{d \to \infty}\psi_{d,p,F}.$$
 Finally, since $p \in [0,1]$, we can average the above over $p$, obtaining 
 a single number, to be called a \emph{rule decompression ratio}, defined as
 $$\psi_{F}=\frac{\int_0^1 \psi_{p,F} \, dp} {\int_0^1 dp }=\int_0^1 \psi_{p,F} \, dp.$$
 
 Computing $\psi_{F}$ for a given CA is not easy, although we will see later that it can be accomplished in some cases. We can, however, compute its approximate value
 as follows.
 
 For a given $p \in [0,1]$ and $d$, we set $n=pd$ 
 and randomly generate a binary string of length $d-2$ with exactly
 $n-2$ ones, selecting it randomly with uniform probability from the set of
 all such strings. We add 1 in front and at the end of this string, obtaining a string
 with $n$ ones having diameter $d$. We then add a number of zeros in front and end obtaining a finite configuration of length $M$. 
 We then iterate  rule $F$ for a sufficient number of time steps so that the diameter
  expands to its limiting value and compute $\psi_{d,p,F}$. This is repeated for values of $p$
  ranging from 0 to 1, with small increments.
    We then plot the decompression ratio versus $p$. This, if $d$ is sufficiently large, approximates the graph of $\psi_{p,F}$ vs. $p$.

\begin{figure}
 \begin{center}
  \includegraphics[width=12cm]{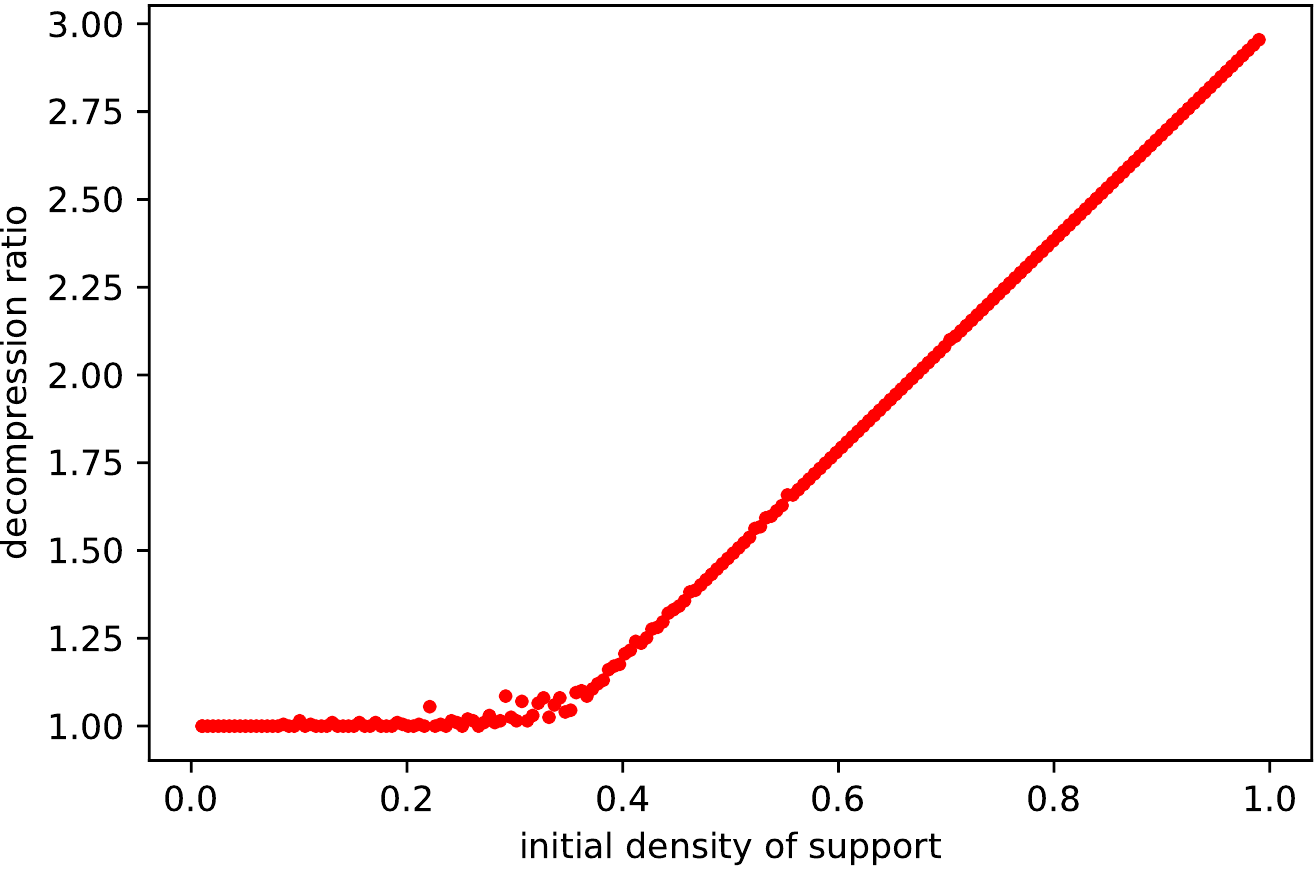}
  \end{center}
  \caption{Decompression ratio $\psi_{d,p,F}$ vs. initial density of support $p$
  for rule 60200. Graph obtained numerically by generating random samples of configurations with 
  support  diameter $d=200$ and iterating rule 60200 until
  diameter stops changing.}\label{fig1}
 \end{figure}

 Figure~\ref{fig1} shows an example of such a plot for rule 60200.
  From this graph, we could conjecture that for rule 60200
  the decompression factor is
  $$\psi_{p,F} = \begin{cases} 1 &\mbox{if } p < 1/3, \\
  3p & \mbox{otherwise.}  \end{cases}$$
With the above, we can now compute  the rule decompression ratio for rule 60200,  
  $$\psi_{F}=\int_0^{1/3} dp+\int_{1/3}^{1} 3p\,dp =\frac{1}{3}+\frac{4}{3}=\frac{5}{3}.$$
For other number conserving rules the rule decompression ratio can be obtained in a similar way. We will carry out this procedure for rules with 3, 4 and 5 inputs in order to fin the ``best'' rule among them, that is, the rule which features the largest decompression ratio.
  \section{Rules with three inputs}
Among the number conserving rules with three inputs, 
three are trivial. These are rules 170 and 240 (shifts) as well as rule 204 (identity). All three preserve diameters of configurations with finite support, thus for them $\psi_F=1$.

Rule 184 (and its reflected version rule 226) is more interesting.
 Due to the nature of rule 
184, for a given $ x \in D_{d,n}$,
the diameter of $F^k(x)$ is non-decreasing with $k$,
and reaches the limiting value after the final number of steps. Define, therefore, the \emph{final diameter} as
$$\lim_{k \to \infty} \diam F^k(x).$$

\begin{figure} 
 \begin{center}
  (a)\includegraphics[width=7cm]{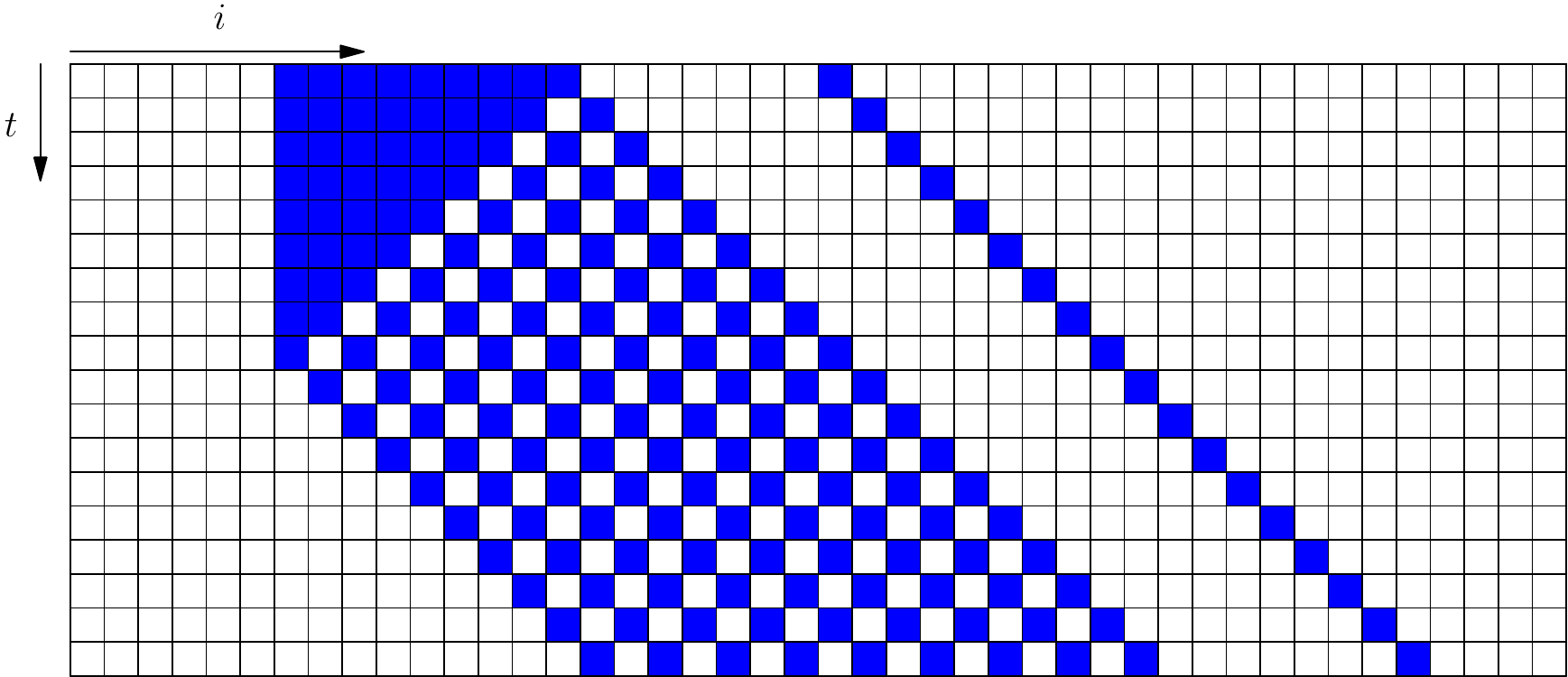}\includegraphics[width=7cm]{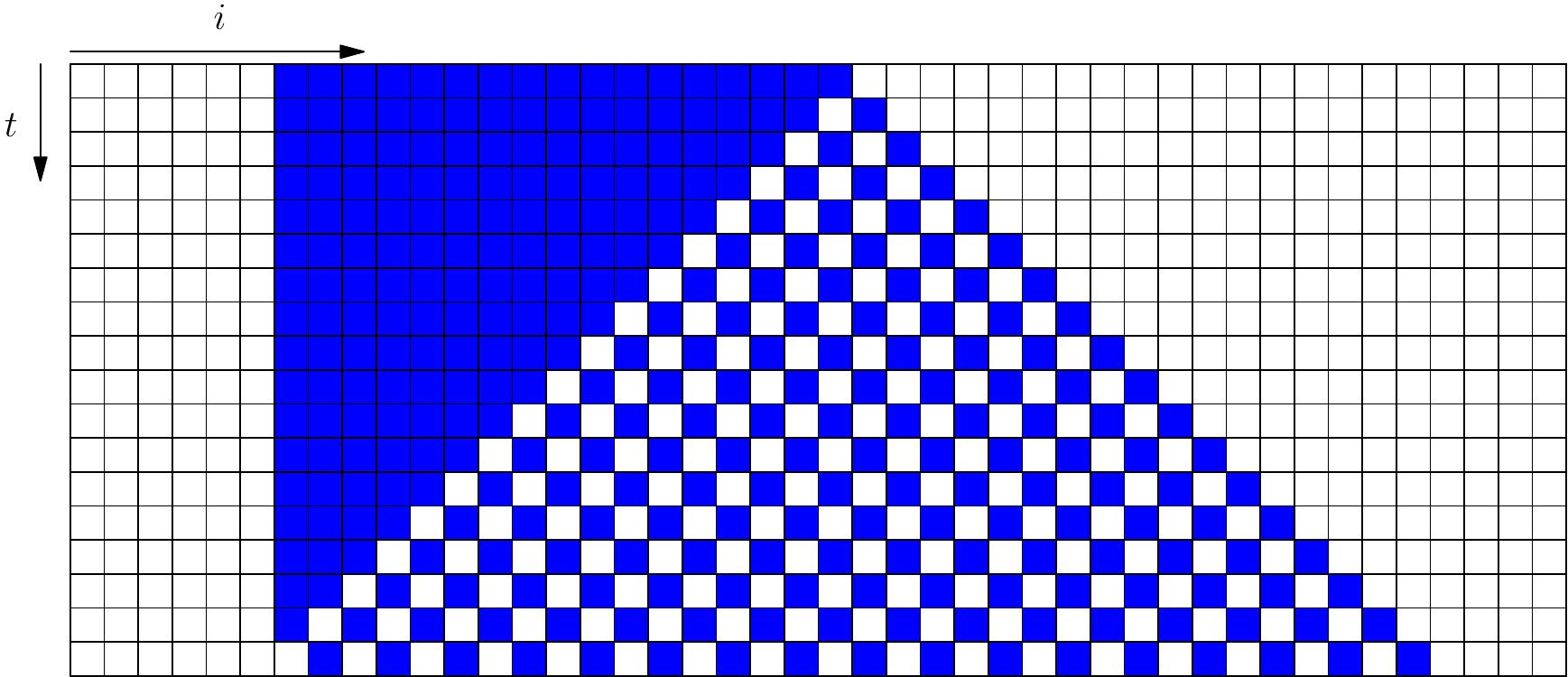}\,\,(b)\\
   (c)\includegraphics[width=7cm]{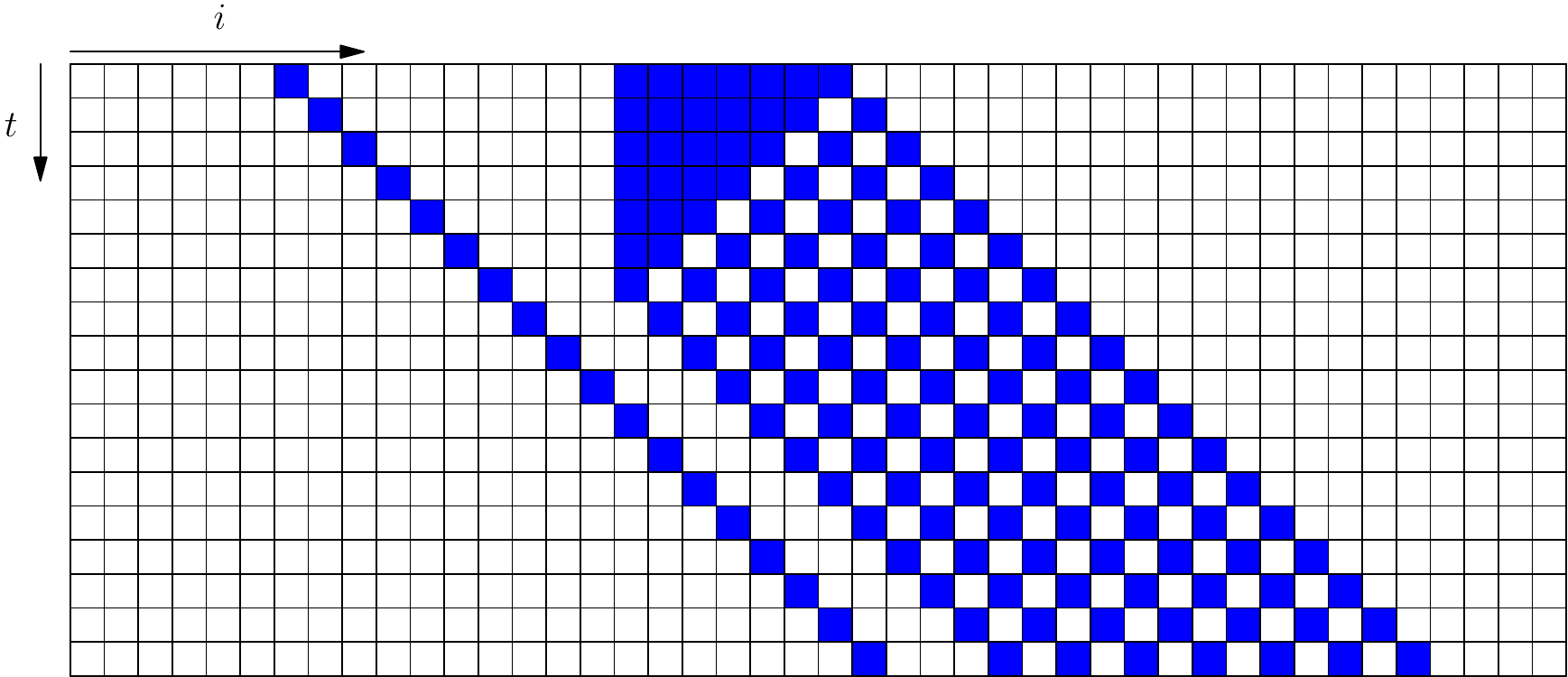}\includegraphics[width=7cm]{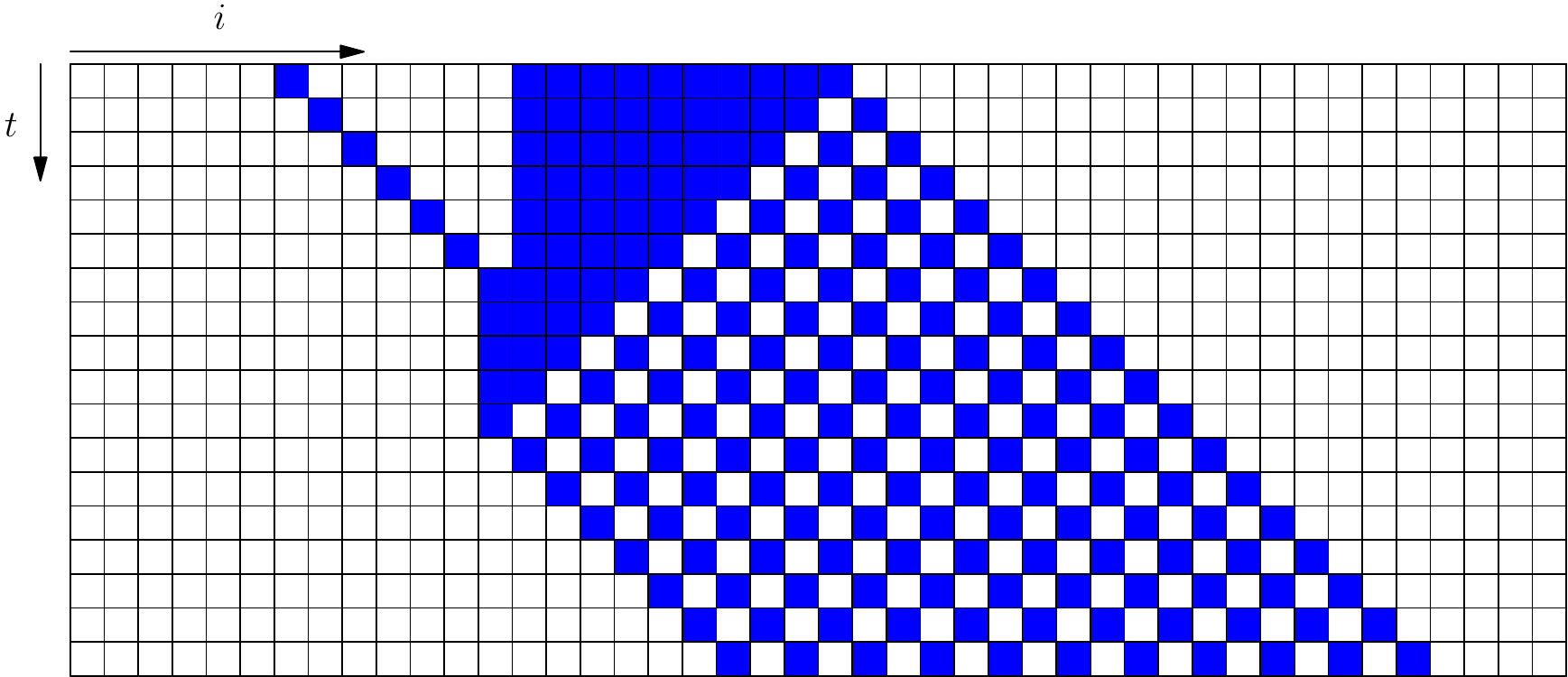}\,\,(d)\\
   (e)\includegraphics[width=7cm]{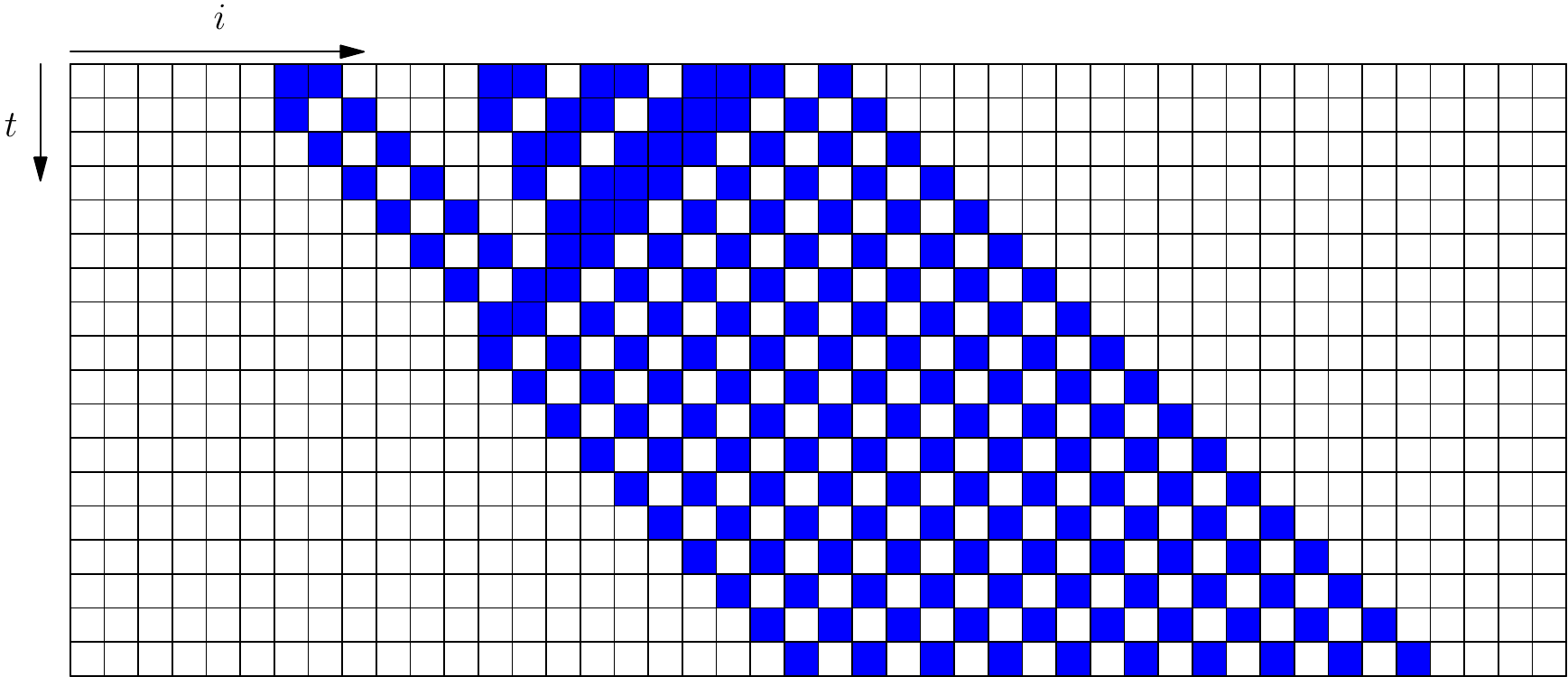}\includegraphics[width=7cm]{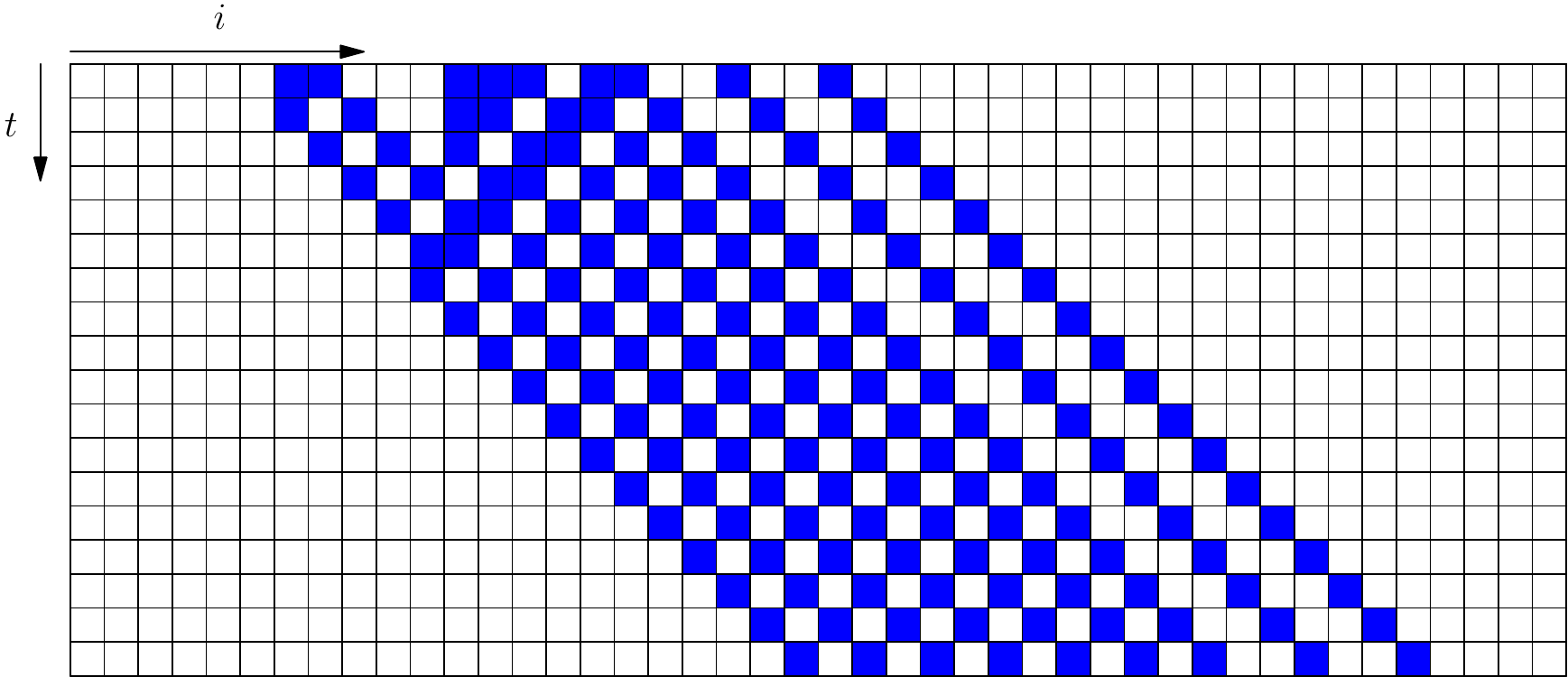}(f)\,\
 \end{center}
\caption{Spatiotemporal patterns for rule 184 used in the proof of Proposition \ref{propbounds}.}\label{r18pats}
\end{figure}

For a given $x \in D_{d,n}$, the final diameter is bounded. One can show the following.
\begin{proposition}\label{propbounds}
Let $d>2$, $2 \leq n \leq d$. For rule 184, for every $x \in D_{d,n}$, the final diameter of $x$
is bounded by
 \begin{equation}
 d_{min} \leq \lim_{k \to \infty} \diam F^k(x) \leq d_{max},
 \end{equation}
 where
 \begin{equation}\label{dmin}
 d_{min} = \begin{cases} d &\mbox{if } 2 \leq n \leq d/2, \\
 2n-1 & \mbox{otherwise,}  \end{cases}
 \end{equation}
 and
 \begin{equation}\label{dmax}
d_{max} = \begin{cases} d+n-2 &\mbox{if } n<d, \\
 2d-1 & \mbox{if $n=d$.}  \end{cases}
 \end{equation}
 \end{proposition}
 \emph{Proof.} Let us consider $d_{max}$ first. Rule 184
 can be viewed as particle system, where each particle (site in state 1)
 will move to the right in the next time step if its right neighbour is empty \cite{paper8}. Since the rightmost particle has always 0 as its right neighbour, it will always move. The rightmost particle, on the other hand, can be stopped,
 and if it is stopped, diameter grows (because the rightmost one moves at the same time). Therefore, in order to maximized the final diameter, we need to
 put all inside particles grouped in a solid block on the left,
 as shown in Figure~\ref{r18pats}a.
 If we do it, the leftmost site will be stopped for
 $n-2$ steps, so the diameter will increase by $n-2$, reaching final value $d+n-2$.
 The only exception to this is the case when $n=d$, as shown in  Figure~\ref{r18pats}b. In this case, the leftmost site is stopped for one more extra step, so that
 the final diameter is $d+n-2 +1=2d-1$. This yields the upper bound given in eq. (\ref{dmax}).

In order to achieve the minimal final diameter, we need to group inside particles
as far to the right as possible, as shown in Figures~\ref{r18pats}c and \ref{r18pats}d. If $n \leq d/2$, the leftmost site is never stopped, so that the diameter does not change at all and is always equal to $d$, as in Figure~\ref{r18pats}c.
If $n>d/2$, the final pattern is of the form $101010\ldots10$, that is,
it consists of $n$ repeated pairs $10$, as shown in in Figure~\ref{r18pats}d.
Diameter of this configuration is $2n-1$. This yields the lower bound given in eq. (\ref{dmin}).

\begin{proposition}
Let $d>2$, $2 \leq n \leq d$. The number of elements
of $D_{d,n}$ having the final diameter $d_f$,
to be denoted as $N(d_f)$, is
given by
\begin{equation}\label{nrofblocksformula}
N(d_f)
 = \begin{cases} 
 \displaystyle \binom{d-2}{d_f-n}-\binom{d-2}{d_f-n+1}
 & \mbox{if } d_f\neq2n-1,\\[1.5em]
 \displaystyle \binom{d-2}{n-2}-\binom{d-2}{n}
 &\mbox{if } d_f=2n-1. 
  \end{cases}
 \end{equation}
\end{proposition}
\emph{Proof.}
We will start with the case $d_f=2n-1$. As we already know from the proof of the
previous proposition, $d_f=2n-1$ means that the final pattern 
consists of alternating 1's and 0's, as in Figures ~\ref{r18pats}d and
~\ref{r18pats}e. In order to reach this configuration, we must make sure that in the
initial configuration any cluster of zeros disappears before the final
configuration is reached. Let us define the rightmost substring of
a given string $s$ as the substring of $s$ which ends at the end of $s$.

The final configuration $101010\ldots1$ will be obtained if any rightmost substring of the support of the initial configuration has no more 0's than 1's.
The number of binary strings of length $d$ staring and ending with 1, having exactly
$n$ 1's and having the property that any rightmost substring of it has no more 0's than 1's is equal to 
$$
 \binom{d-2}{n-2}-\binom{d-2}{n},
$$
exactly as claimed in eq. (\ref{nrofblocksformula}). 

When $d_f\neq2n-1$, the final configuration
will not be  $101010\ldots1$, but will include some extra 0's,
as in the example in Figure~\ref{r18pats}f.
The number of such extra zeros is $\epsilon=d_f-(2n-1)$.
Final configuration with $\epsilon$ extra zeros is obtained when  among the rightmost substrings of the support of the initial configuration the maximal excess of 0's over 1's is  $\epsilon$. The number of binary strings of length $d$ staring and ending with 1, having exactly
$n$ 1's and having the property that among the rightmost substrings of each of of them the maximal excess of  0's over 1's equals $\epsilon$ is given by
$$
 \binom{d-2}{\epsilon+n-1}-\binom{d-2}{\epsilon+n}
 =\binom{d-2}{d_f-n}-\binom{d-2}{d_f-n+1},
$$
again in agreement with eq. (\ref{nrofblocksformula}).
$\Box$

We can now compute the decompression ratio for rule 184 as
$$
 \psi_{d,p,F}
 =
 \frac{ \sum_{d_f=d_{min}}^{d_{max}} \frac{d_f}{d} N(d_f)}
 { \sum_{d_f=d_{min}}^{d_{max}} N(d_f)}.
$$
The denominator of the above,
as we already remarked, equals to 
$\binom{d-2}{n-2}$, thus we obtain
\begin{equation}\label{r184decomformula}
 \psi_{d,p,F}
 =\frac{1}{d} {\binom{d-2}{n-2}}^{-1}
  \sum_{d_f=d_{min}}^{d_{max}} d_f N(d_f),
\end{equation}
where $n=pd$ is rounded to the nearest
integer, $N(d_f)$ is given by eq. (\ref{nrofblocksformula}),
and $d_{min}$, $d_{max}$ are given by eqs. (\ref{dmin}--\ref{dmax}).

\begin{figure}
 \begin{center}
  \includegraphics[width=12cm]{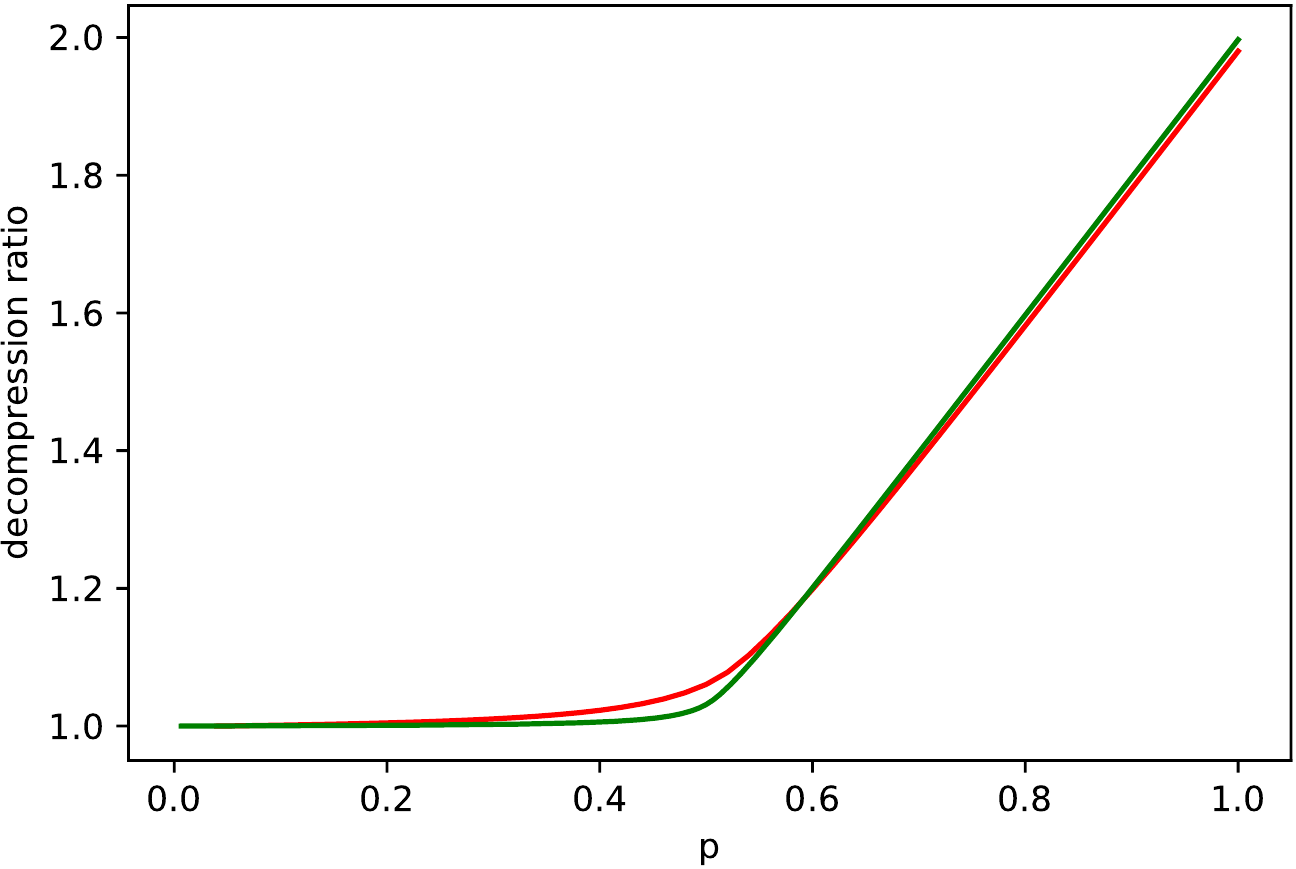}
 \end{center}
\caption{Graph of the decompression ratio $ \psi_{d,p,F}$  as a function of  $p$ for rule 184, 
obtained from eq. (\ref{r184decomformula}) using $d=50$ (red) and $d=300$ (green).}\label{r18rtheoretical}
\end{figure}

Although it does not seem to be possible to compute the closed
form of the sum in the above formula, numerical
plot of the decompression ratio can easily be produced.
Figure~\ref{r18rtheoretical} shows the 
plot of  $ \psi_{d,p,F}$  as a function of  $p$ 
for two different values of $d$.
One can see that as $d$ increases, the graph
develops a ``sharp corner'' at $p=1/2$.
Indeed, our  numerical investigations of the behaviour of  $ \psi_{d,p,F}$ for large $d$ indicate that  for rule 184,
 \begin{equation}
 \psi_{p,F}=\lim_{d \to \infty}\psi_{d,p,F} = \begin{cases} 1 &\mbox{if } p < 1/2, \\
 2p & \mbox{otherwise.}  \end{cases}
 \end{equation}
This yields $\psi_F=\int_0^1 \psi_{p,F}\,dp=1.25$.
 \section{Four input rules}
 In section 2 we remarked that there are 22 4-input rules.
 Some of these are not really 4-input, but they  effectively depend on 3 inputs (or less), and as such they do not need to be considered because they were discussed in the previous section.
 What is left is 14 rules falling into
 7 equivalence classes with respect
 to spatial reflection,
$ \{43944, 65026\}$, 
$\{48268, 63544\}$, 
$\{48770, 60200\}$,      
$\{49024, 59946\}$, 
$\{51448, 62660\}$,  
$\{52930, 58336\}$, and 
$\{56528, 57580\}$. It is obvious that
each pair has the same rule decompression ratio and the same graph of $\psi_{d,p,f}$ vs. $p$, thus only one of each needs to be considered.

One should remark at this point that it is customary in CA research to divide rules into equivalence classes
with respect to the group generated by spatial reflection and conjugation.
For the 14 aforementioned 4-input rules these equivalence classes
would be 
$\{48770, 60200\}$, 
$\{43944, 49024$, $59946, 65026\}$, 
$\{48268, 52930, 58336, 63544\}$,
and $\{51448, 56528, 57580, 62660\}$.
However, since our definition
of the decompression ratio
is not symmetric with respect to
interchange of 0's and 1's, rules
belonging to the same equivalence class of this type may have different graphs of $\psi_{d,p,F}$ and different $\psi_F$. 
This classification, therefore, is not very useful
for our purposes.

\begin{figure}
 \begin{center}
 (a)\includegraphics[width=7cm]{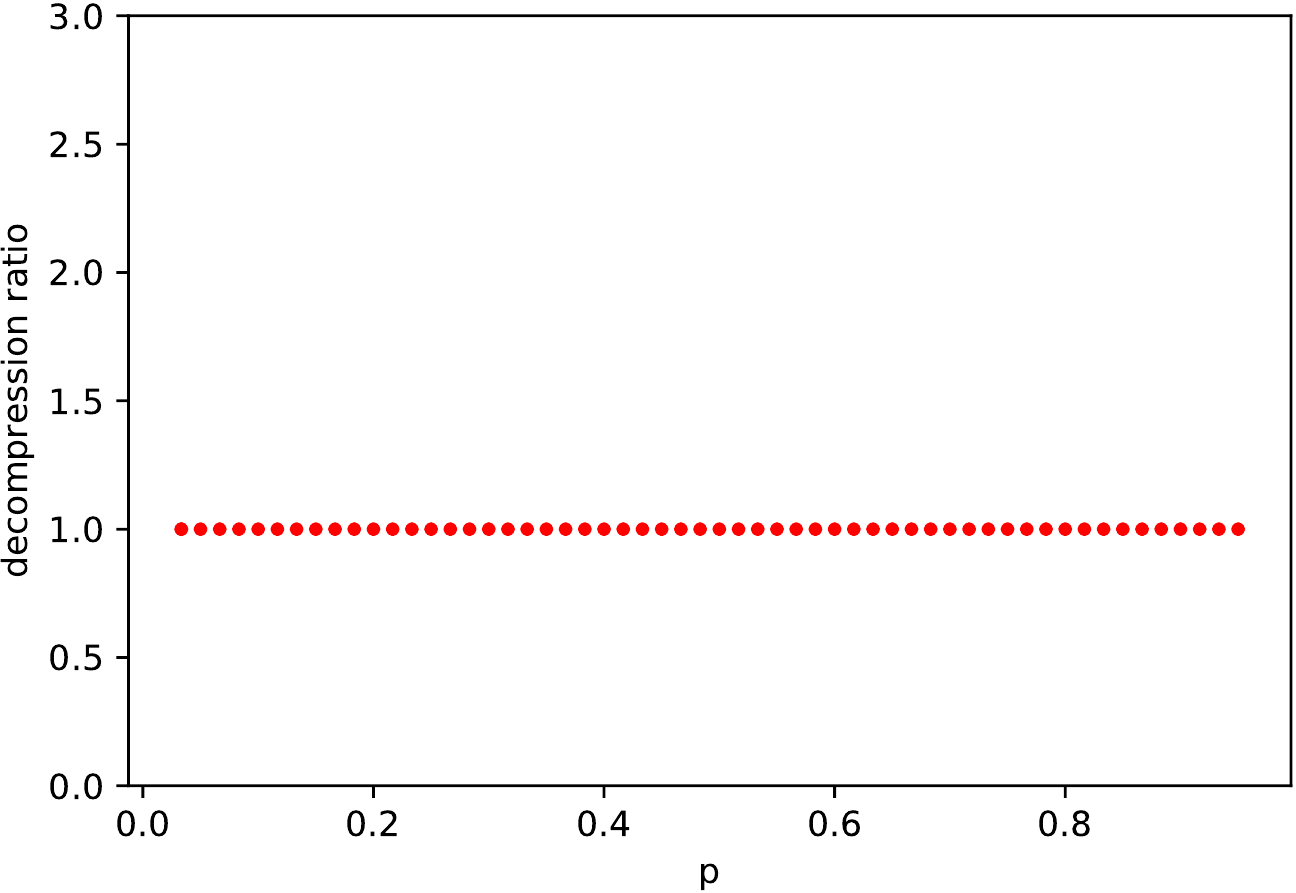}\,\,\,\,\includegraphics[width=7cm]{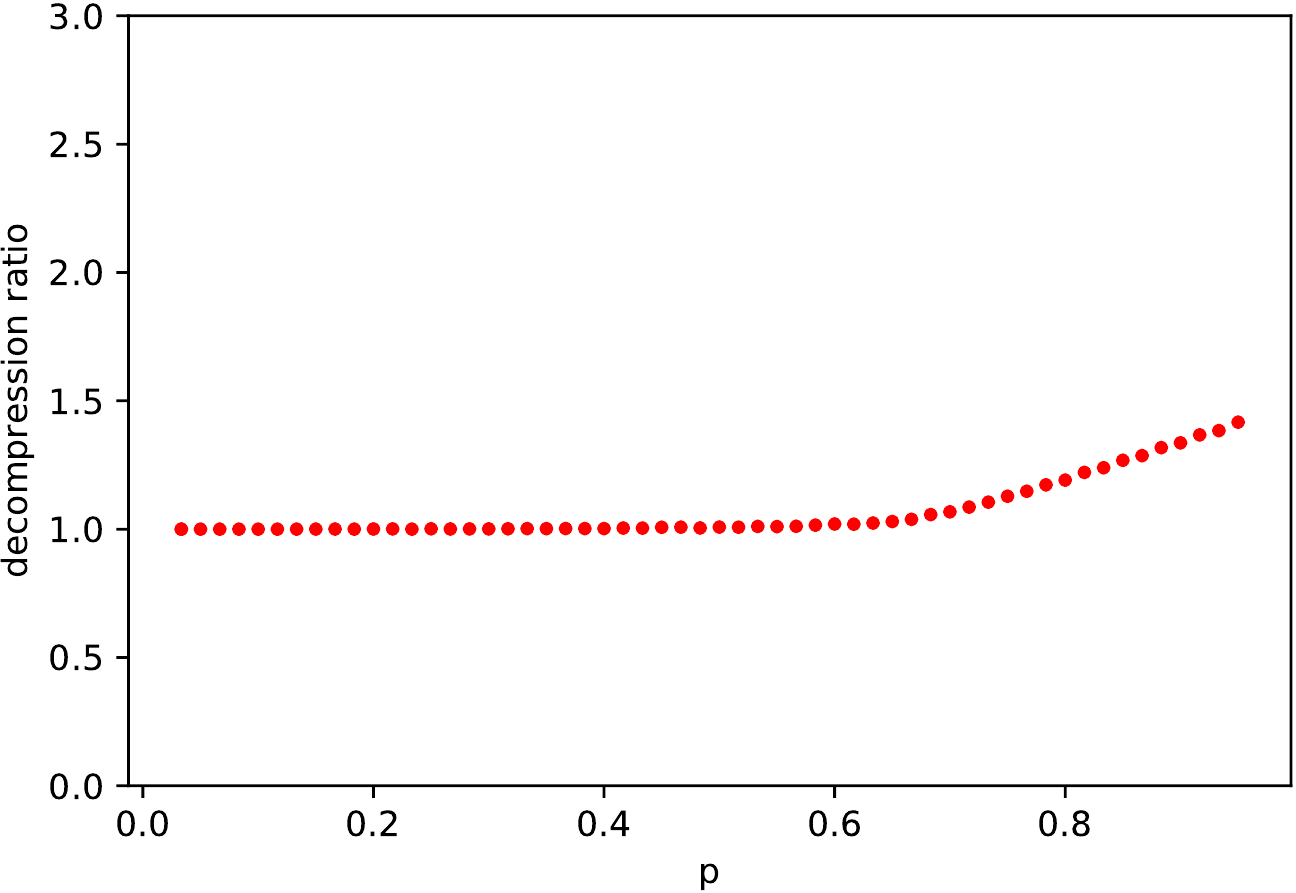}(b)\\  (c)\includegraphics[width=7cm]{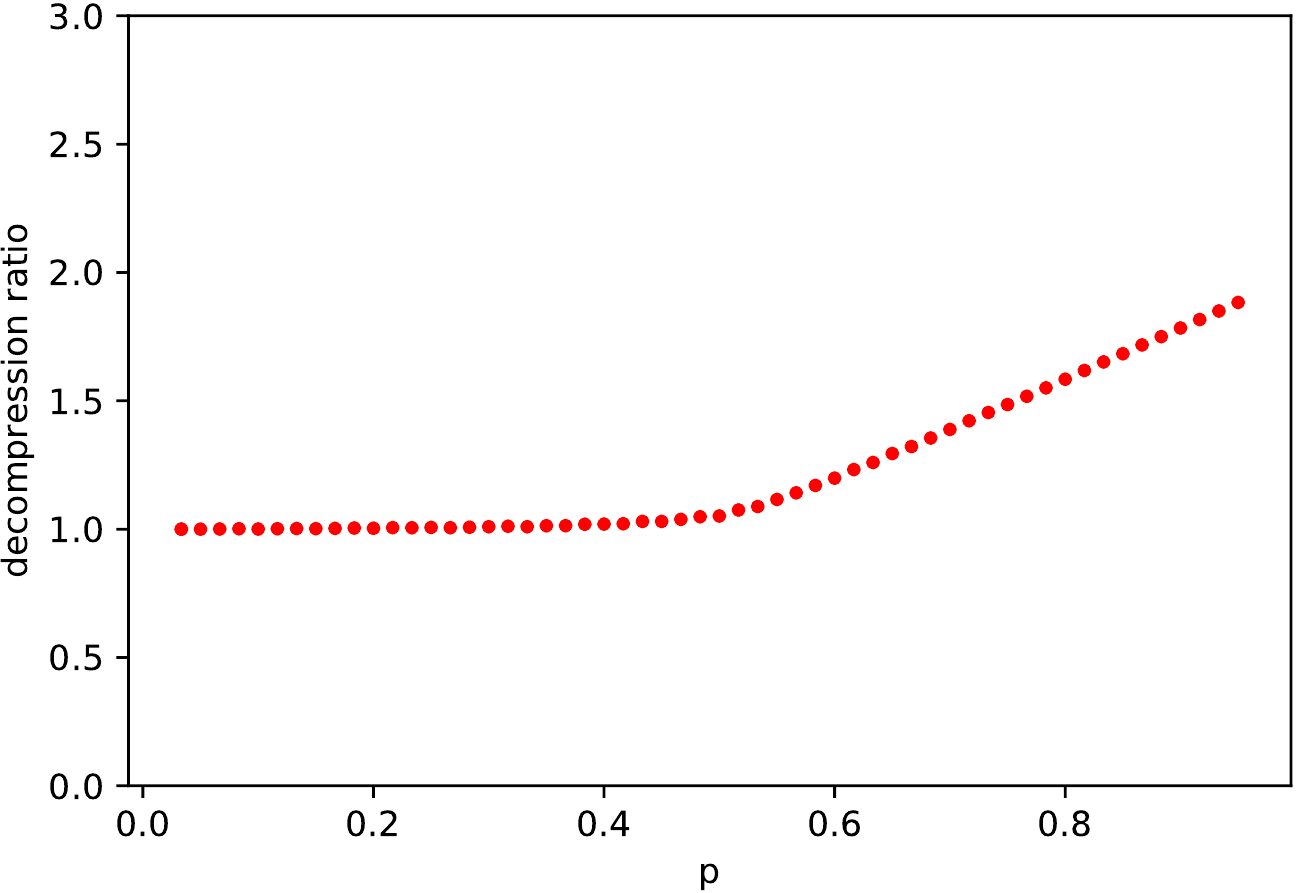}\,\,\,\,
 \includegraphics[width=7cm]{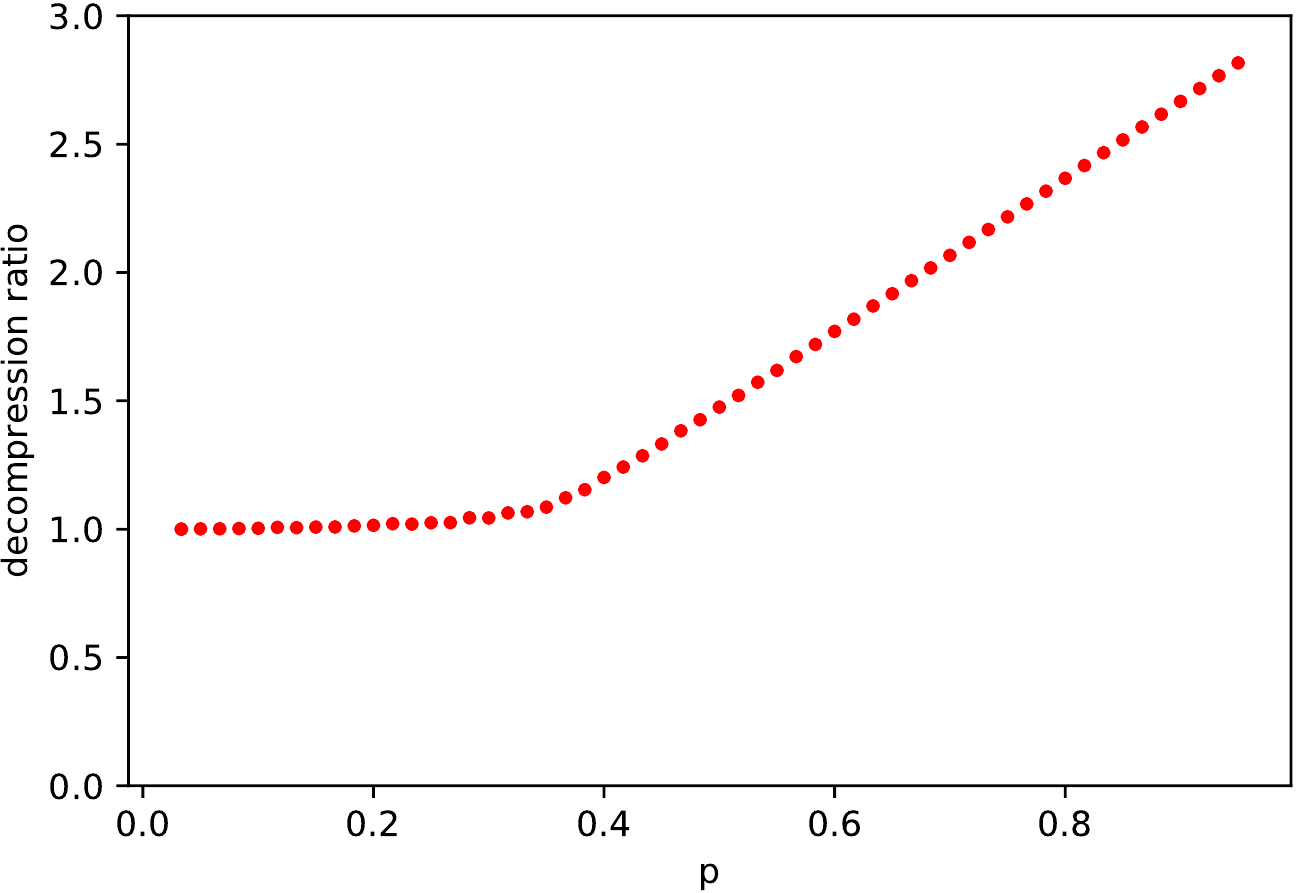}(d) 
 \end{center}
\caption{Four types of graphs of numerically computed decompression ratio $\psi_{d,p,F}$ as a function of $p$ for 4-input rules. Graphs obtained by random sampling of
100 initial configurations for each $p$ using initial diameter $d=60$.
Representative graphs are shown: they vary sightly between rules, but the differences disappear as $d \to \infty$.
}\label{fourinputgraphs}
 \end{figure}

We investigated the behaviour of  of $\psi_{d,p,F}$  numerically for
all representative 4-input rules. It turns out that for rules
$\{51448, 62660\}$ the diameter grows unbounded,
and thus we say that $\psi_F = \infty$ for these two rules.
The remaining rules have finite $\psi_F$, and their graph
of $\psi_{d,p,F}$ vs. $p$ is one of four types shown in Figure~\ref{fourinputgraphs}.
 We found a rather remarkable fact, namely,  for $d\to \infty$, all finite decompression ratios $\psi_{p,F}$ 
 appear to assume the same functional form,
\begin{equation}\label{genform}
 \psi_{p,F}=\lim_{d \to \infty}\psi_{d,p,F} = \begin{cases} 1 &\mbox{if } p < p_c, \\
 p/p_c & \mbox{otherwise,}  \end{cases}
 \end{equation}
 where $p_c$ takes one of the values $1/3$, $1/2$, $2/3$ or $1$.
 The graph types as well as $p_c$ for all 4-input rules
 are shown in Table~\ref{3an4rinputtable} (for completeness,  3-input rules are included as well).
 
 Using eq. (\ref{genform}) it is now straightforward to compute the rule decompression
 ratio,
 \begin{equation}
  \psi_F= \int_0^1 \psi_{p,F} \, dp= \frac{1+p_c^2}{2 p_c},
 \end{equation}
and the relevant values are shown in the last column of Table~\ref{3an4rinputtable}.
 \begin{table} 
\begin{center}
\begin{tabular}{|l|l|c|l|l|l|}\hline
& $W(f)$ & graph type & $p_c$ & $\psi_F$& decimal value of $\psi_F$  \\ \hline
3-input&170,240        & a & 1   & 1       & 1.0\\
       &204            & a & 1   & 1       & 1.0 \\
       &184,226        & c & 1/2 & 5/4     & 1.25 \\ \hline
4-input& 43944, 65026  & d & 1/3 & 5/3     & 1.666\ldots \\
       & 48268, 63544  & c & 1/2 & 5/4     & 1.25 \\
       & 48770, 60200  & d & 1/3 & 5/3     & 1.666\ldots \\
       & 49024, 59946  & b & 2/3 & 13/12   & 1.0833\ldots \\ 
       & 51448, 62660  & n/a &n/a& $\infty$& n/a\\
       & 52930, 58336  & d & 1/3 & 5/3     & 1.666\ldots \\
       & 56528, 57580  & a & 1   & 1       & 1.0 \\
         \hline
\end{tabular}
\end{center}
\caption{Estimated rule decompression ratio 
for 3-input and 4-input rules.}\label{3an4rinputtable}
\end{table}
 
From the graphs and table we can conclude that the highest finite 
rule decompression ratio obtainable for 4-input rules is $5/3$.
This is are only slightly better
than the value for rule 184. None of the 4-input rules are particularly
good for decompressing initial configurations with small densities:
in fact, they do not decompress at all if the density of support of the initial configuration is less than $1/3$. When the density increases beyond the critical value $p_c$, the 
decompression ratio grows linearly, reaching the maximum value for strings consisting of only 1's.

\begin{figure}
 \begin{center}
  \includegraphics[width=12cm]{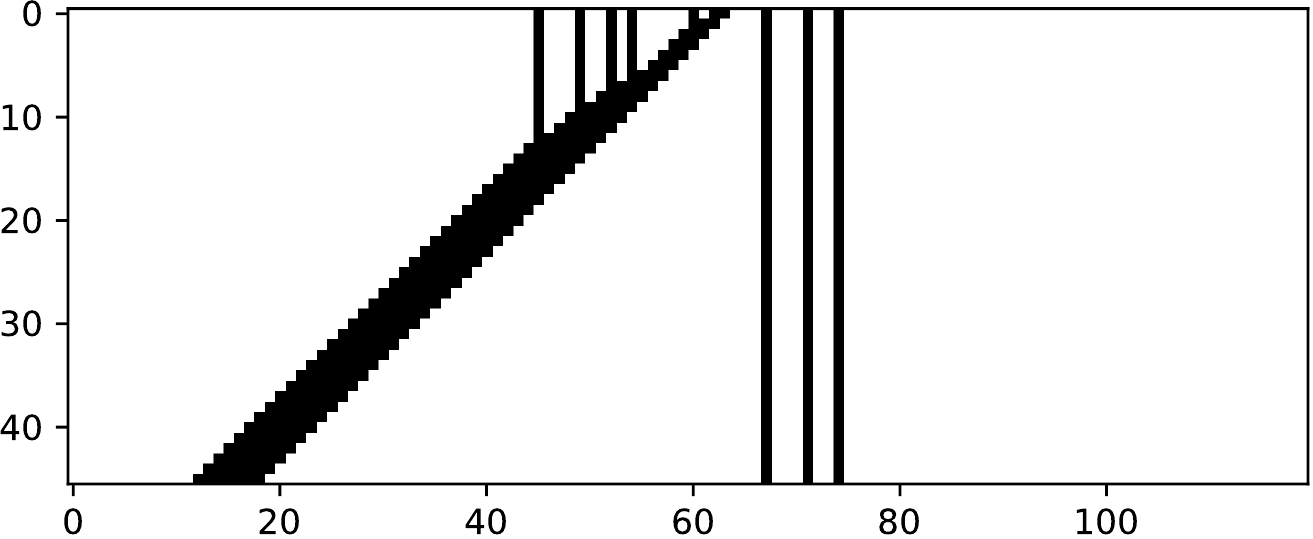}
 \end{center}
\caption{Example of a spatiotemporal pattern for rule 51448 demonstrating
unbounded growth of the diameter.}\label{unbounded}
\end{figure}

Rule 51448, for which the decompression ratio is infinite, indeed expands diameters 
without any bound, yet it is not performing ``spreading'' of particles in a particularly
satisfactory way. In fact, as we can see in Figure~\ref{unbounded}, it performs consolidation
of some particles on the right, which then move as a solid block to the right while other particles remain in place.
This rule does not increase disorder, or, in other words, does not
increase entropy in the same way as diffusion proces illustrated in Figure~\ref{randif}.
\section{Five input rules}
There exist 428 number conserving CA rules with 5 inputs. They can be divided into
215 equivalence classes with respect to spatial reflection, or into 129 classes
with respect to reflection and conjugation. In order to survey the types of behaviour
which is possible in 5-input rules and keep their number manageable, we decided to use representatives of the later 129 classes. The representative chosen for each equivalence class is the rule with
with minimal Wolfram number. One should keep in mind, however, as remarked in the previous section, 
that such a representative
may not necessarily have the same decompression ratio as other members of the same class.
Out of the aforementioned 129 rules there are 13 which effectively depend on 4 inputs or less,
and these are eliminated. This leaves 116 rules.

\begin{figure}
\begin{center}
 \includegraphics[width=6cm]{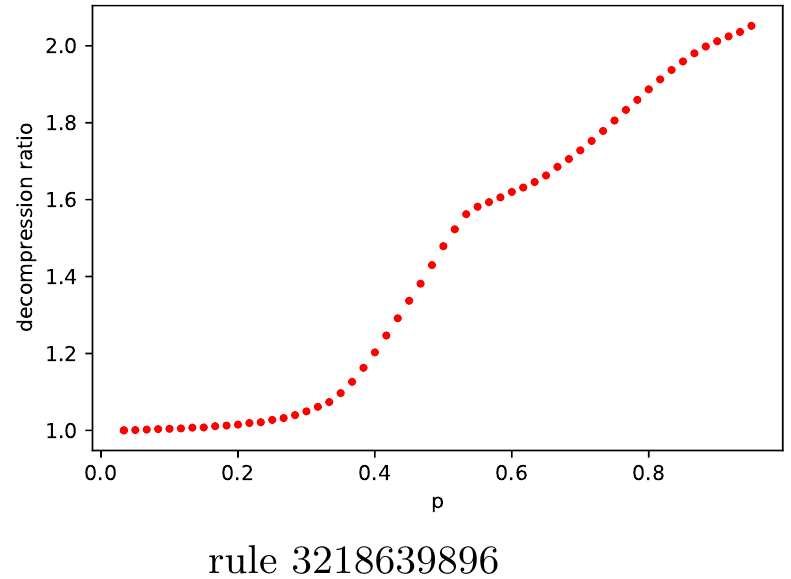}
 \includegraphics[width=6cm]{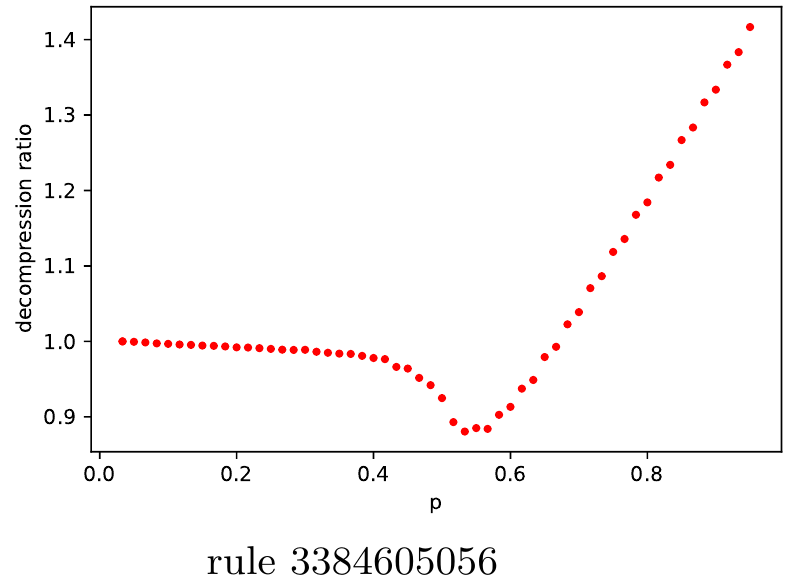}\\[1em]
 \includegraphics[width=6cm]{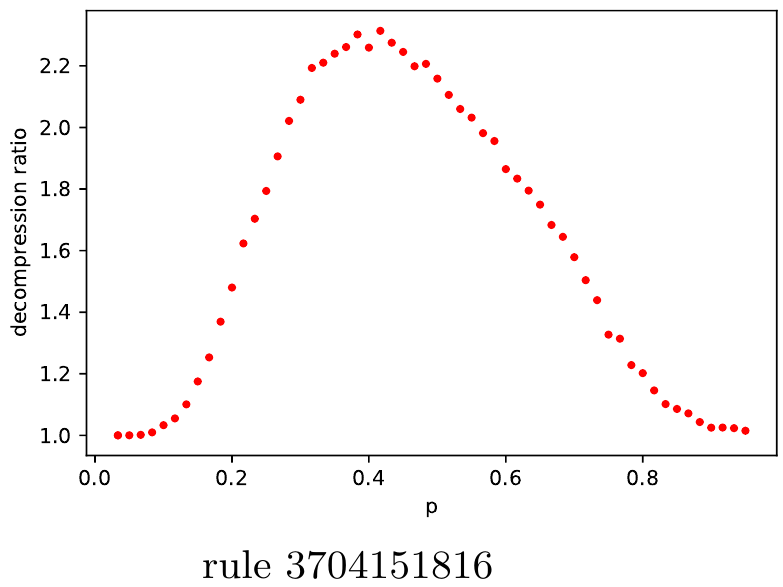}
 \includegraphics[width=6cm]{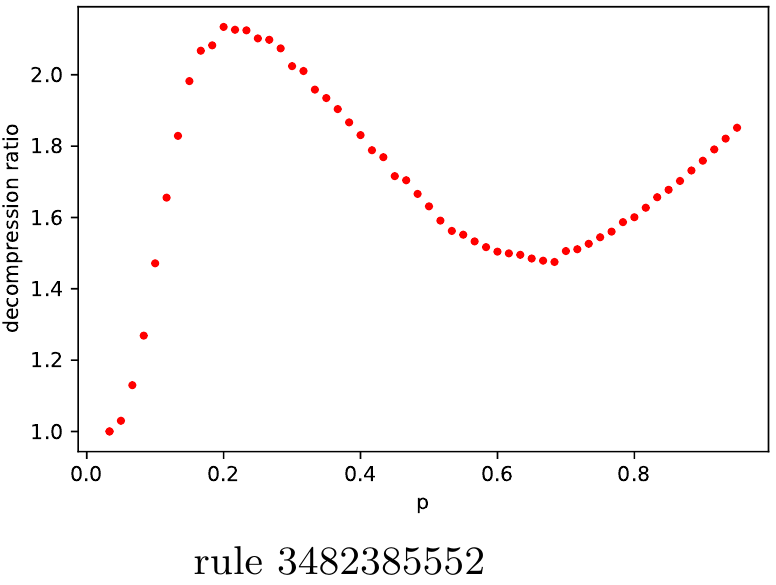}
 \end{center}
\caption{Examples of 5-input rules with ``irregular'' decompression graphs.}\label{irreg}
\end{figure}

We found that there are three types of behaviour of $\psi_{d,p,F}$ as a function of $p$:
\begin{enumerate}
 \item rules exhibiting infinite decompression ratio. We found 5 rules of this type.
 \item rules with ``hockey-stick'' type graphs of the form of eq. (\ref{genform}). There are 85 rules of this type.
 \item 26 rules with ``irregular'' decompression graphs, as shown in examples of
 Fig.~\ref{irreg}.
\end{enumerate}

Regarding the second type, the smallest
value of $p_c$ we found is $p_c=1/4$.
This, together with the results for 
rules with 3 and 4 inputs presented earlier, suggests the following  conjecture.
\begin{conjecture}
If a CA rule with $n$ inputs exhibits decompression ratio
of the form of eq. (\ref{genform}),
then $$p_c \geq\frac{1}{n-1}.$$
\end{conjecture}
We have no proof of the above, but some intuition can be offered.
A given site can interact with $n-1$ other sites, due to the nature of the local function. The expansion of a pattern can only happen if close particles are repelling each other. Obviously,
they can repell each other only if they are located within
the range of interaction. If we have on average less than 
one particle per $n-1$ sites, they will not be able to interact,
thus there will be no expansion, and $\phi_{p,F}$ will remain 1.
This means that $p_c=1/(1-n)$ is the lowest threshold for
the transition from the non-expanding to the expanding regime.

As for the last type, rules with ``irregular'' decompression graphs, 
these often are very slowly converging. In fact, some rules which we now 
consider type 2, in our initial experiments were classified as type 3.
Only when we increased the diameter and the number of iterations
they assumed the shape of the ``hockey stick''. For this reason,
it is quite possible that future more extensive numerical 
experiments may reveal that some of type 3 rules still exhibit the form of eq.~(\ref{genform}).

\section{Conclusions}
We found that deterministic number conserving cellular automata rules exhibit mostly very limited resemblance to diffusion. 
Although they expand the patterns
with finite support, the vast majority
of them spread only patterns with
high density, and not patterns
with low density. It is rather remarkable that the large percentage of
them exhibits decompression ratio graphs
in the form of of eq. (\ref{genform}).
The existence of the transition
at $p=p_c$ and the fact that such transition occurs in a large number
of rules needs to be investigated further.

There exist, however, a small number of rules which exhibit infinite decompression ratio. These expand diameters of
patterns with finite support, including
even patterns with low density. Their behavior, however, differs from the random diffusion because they only increase the diameter without increasing the
spatial entropy of the configuration. 
In order to construct a CA rule which is 
a better model of random diffusion 
one  will likely need more than
two states, with some states
acting like pseudo-random generators, while others contributing to the increase of the diameter. Work in this direction is ongoing, and results will be reported elsewhere.


\begin{thebibliography}{1}

\bibitem{b93}
N.~Boccara.
\newblock Transformations of one-dimensional cellular automaton rules by
  translation-invariant local surjective mappings.
\newblock {\em Physica D}, 68:416--426, 1993.

\bibitem{paper8}
N.~Boccara and H.~Fuk{\'s}.
\newblock Cellular automaton rules conserving the number of active sites.
\newblock {\em J. Phys. A: Math. Gen.}, 31:6007--6018, 1998.

\bibitem{paper12}
N.~Boccara and H.~Fuk{\'s}.
\newblock Number-conserving cellular automaton rules.
\newblock {\em Fundamenta Informaticae}, 52:1--13, 2002.

\bibitem{Durand2003}
B.~Durand, E.~Formenti, and Z.~R{\'o}ka.
\newblock Number-conserving cellular automata {I}: decidability.
\newblock {\em Theoretical Computer Science}, 299:523--535, 2003.

\bibitem{Formenti2003}
E.~Formenti and A.~Grange.
\newblock Number conserving cellular automata {II}: dynamics.
\newblock {\em Theoretical Computer Science}, 304:269--290, 2003.

\bibitem{Hattori91}
T.~Hattori and S.~Takesue.
\newblock Additive conserved quantities in discrete-time lattice dynamical
  systems.
\newblock {\em Physica D}, 49:295--322, 1991.

\bibitem{Moreira03}
A.~Moreira.
\newblock Universality and decidability of number-conserving cellular automata.
\newblock {\em Theor. Comput. Sci.}, 292:711--721, 2003.

\bibitem{Pivato02}
M.~Pivato.
\newblock Conservation laws in cellular automata.
\newblock {\em Nonlinearity}, 15:1781--1793, 2002.

\bibitem{Wolfram94}
S.~Wolfram.
\newblock {\em Cellular Automata and Complexity: Collected Papers}.
\newblock Addison-Wesley, Reading, Mass., 1994.

\end{thebibliography}
\end{document}